\documentclass{aastex62}

\received{...}
\revised{...}
\accepted{...}
\submitjournal{ApJ}

\shorttitle{Periodic Nova in M31}
\shortauthors{Marelli et al.}

\begin{document}

\title{The first orbital period of a very bright and fast Nova in M31: M31N 2013-01b}

\correspondingauthor{Martino Marelli}
\email{marelli@lambrate.inaf.it}

\author{Martino Marelli}
\affil{INAF - Istituto di Astrofisica Spaziale e Fisica Cosmica Milano, via E. Bassini 15, I-20133 Milano, Italy}

\author{Domitilla De Martino}
\affiliation{INAF - Osservatorio Astronomico di Capodimonte, Salita Moiariello 16, I-80131 Napoli, Italy}

\author{Sandro Mereghetti}
\affiliation{INAF - Istituto di Astrofisica Spaziale e Fisica Cosmica Milano, via E. Bassini 15, I-20133 Milano, Italy}

\author{Andrea De Luca}
\affiliation{INAF - Istituto di Astrofisica Spaziale e Fisica Cosmica Milano, via E. Bassini 15, I-20133 Milano, Italy}

\author{Ruben Salvaterra}
\affiliation{INAF - Istituto di Astrofisica Spaziale e Fisica Cosmica Milano, via E. Bassini 15, I-20133 Milano, Italy}

\author{Lara Sidoli}
\affiliation{INAF - Istituto di Astrofisica Spaziale e Fisica Cosmica Milano, via E. Bassini 15, I-20133 Milano, Italy}

\author{Gianluca Israel}
\affiliation{INAF - Osservatorio Astronomico di Roma, Via Frascati 33, I-00040 Monteporzio Catone, Italy}

\author{Guillermo Rodriguez}
\affiliation{INAF - Osservatorio Astronomico di Roma, Via Frascati 33, I-00040 Monteporzio Catone, Italy}

\begin{abstract}

We present the first X-ray and UV/optical observations of a very 
bright and fast nova in the disc of M31,  M31N 2013-01b. The nova reached a peak magnitude 
$R\sim$15\,mag and decayed by 2 magnitudes in only 3\,days, making it one of the
brightest and fastest novae ever detected in Andromeda. From archival
multi-band data we have been able to trace its fast evolution down to $U>21$\,mag in less
than two weeks and to uncover for the first time the Super-Soft X-ray phase, whose onset 
occurred 10-30\,days from the optical maximum. 
The X-ray spectrum is consistent with a blackbody with a temperature of $\sim$50 eV and 
emitting radius of $\sim$4$\times 10^{9}$ cm, larger than a white dwarf radius, indicating 
an expanded region.
Its peak X-ray luminosity, 3.5$\times 10^{37}$\,erg\,s$^{-1}$, locates M31N 2013-01b 
among the most luminous novae in M31.\\
We also unambiguously detect a short 1.28$\pm$0.02\,h X-ray periodicity that we 
ascribe to the binary orbital period, possibly due to partial eclipses. This makes 
M31N 2013-01b the first nova in M31 
with an orbital period determined. The short period also makes this nova one of the few 
known below the 2-3\,h orbital period gap.\\
All the observed characteristics strongly indicate that M31N 2013-01b harbours 
a massive white dwarf and a very low-mass companion, consistent with being a nova belonging to the disc population
of the Andromeda Galaxy.

\end{abstract}


\section{Introduction} \label{sec:intro}

Classical Novae (CN) are close binary systems ($\rm P_{orb} \sim$ 1.5 - 10\,h) 
consisting of a late type main sequence or red giant secondary
and an accreting white dwarf (WD) primary that experience an outburst triggered 
by a thermonuclear runaway in the hydrogen-rich accreted material \citep{bod08}. CN thus belong
to the cataclysmic variable (CV) class.
After a maximum (brightening up to 7-16\,mag), 
the optical emission declines due to the receding photosphere at a rate which defines the Nova speed class \citep{pay64}.
Part of the material is expelled at high velocities \citep{sho11}.
The decline time, typically defined as the time needed to decline by 2 magnitudes (t2) was found to be related to the peak maximum \citep{del95}
and to the expansion velocity of the ejected  envelope \citep{del02,sha11,sch11}. Fast novae tend to be brighter and to display high expansion velocities. 

After one to a few weeks, the receding atmosphere towards the hotter inner regions is such that the emission moves to X-rays, making the
Nova a Super-Soft X-ray Source (SSS) with temperature of the order of $\sim$20-80\,eV and
X-ray luminosities $> 10^{36}$ erg s$^{-1}$. This phase is powered by stable hydrogen burning within the part of
the accreted envelope and is observable when the ejected matter becomes optically thin to soft X-rays
\citep{ori01,kra02}.

Few novae during the SSS phase were found to display short 
period X-ray oscillations or quasi-oscillations on time-scales of several tens of sec with relatively short duty cycles \citep{bea08,nes15},
ascribed to $g-mode$ pulsations of the burning envelope, which, if confirmed, have the potential to estimate
the WD mass \citep[see][and references within]{nes15,wol18}.
Oscillations on time-scales of thousands of seconds detected during the SSS phase have been
interpreted as non-radial pulsations \citep{dra03,dob10}
while those found to be coherent signify the rotation period of a magnetic WD \citep{pie11,dob10}.
On the other hand, periodic variations on longer (hours) timescales are
compatible with the binary orbit, showing either eclipses \citep{sal08} or X-ray modulations \citep{hen10,hen14,pag10,bea12}.

Most Milky Way novae are found either in the Galactic Bulge or lying in the Galactic plane, thus 
suffering from strong absorption. Novae in Local Group galaxies provide instead
the best targets to study outburst evolution and especially 
the SSS phase. Thanks to its close distance (776\,pc, \citet{dal12}; 
we will adopt this distance throughout this article) 
and low Galactic extinction \citep{dal12} M31 is an ideal target to study nova populations.
\citet{hen10,hen14} published $\sim$25 novae (out of the $\sim$80 known) in M31 that show the SSS phase, 
covering the period until 2012.
Only two of them, M31N 2006-04a and M31N 2011-11e, possess a candidate X-ray orbital period 
(1.6\,h and 1.3\,h respectively), not confirmed due to the low statistics \citep{hen14,pie11}.

The Exploring the X-ray Transient and variable Sky (EXTraS) project \citep{del16} developed new techniques and
tools to extract and describe the timing behavior of archival X-ray sources detected by {\it XMM-Newton/EPIC} \citep{str01,tur01}.
Thanks to the great improvement of EXTraS with respect to the 3XMM source catalog,
we performed a systematic search for eclipsing and/or dipping sources in the {\it XMM-Newton} data set pointing the M31 Galaxy, finding
significant and periodic dips in the light curve of 3XMM J004401.9+412544.
The X-ray source is located at 
R.A.$=00^{{\rm h}}~44^{{\rm m}}~01.^{{\rm s}}97$, Dec.$=+41^{\circ}~25'~44.5''$ (J2000)\footnote{3XMM-DR8: http://xmmssc.irap.omp.eu/Catalogue/3XMM-DR8/3XMM\_DR8.html}
(1$\sigma$ statistical plus systematic error of 2''), positionally consistent with the optical nova
PNV J00440207+4125442=M31N 2013-01b (henceforth M31N2013),
{located at R.A.$=00^{{\rm h}}~44^{{\rm m}}~02.^{{\rm s}}09$, Dec.$=+41^{\circ}~25'~44.4''$ (J2000) (5'' statistical plus systematic error) \citep{hor13}.}
It was detected only once by {\it XMM-Newton} (out of 12 observations)
on Feb.8, 2013, $\sim$14 days after the optical maximum reported by \citet{hor13} and during a follow-up observation of the nova, 
thus confirming the association of the two sources.\\
M31N2013 was found as a bright optical transient \citep{hor13}, with a peak
R-band magnitude on 2013, Jan. 25.728\,UT (this will be considered as T0 for the rest of the paper) of 15.05(7).
Follow-up observations\footnote{IAU CBAT: http://www.cbat..eps.harvard.edu/unconf/followups/J00440207+4125442.html}
showed a rapid fading in a few days. M31N2013 was spectroscopically confirmed
as a He/N or hybrid (Fe\,IIb) nova by \citet{sha13}, displaying broad (FWHM $\rm \sim 5500\,km\,s^{-1}$ Balmer emission lines.
The magnitude and X-ray luminosity (see Table 1) are consistent with a 
nova in M31, while excluding a foreground nova or a nova in our Galaxy
(that would be much more or much less luminous than the other observed novae, respectively).
It is located in the M31 disc \citep{hor13}, further supporting the M31 membership.

During the optical decline, apart from {\it XMM-Newton}, M31N2013 was also observed with the {\it Neil Gehrels Swift} satellite (henceforth {\it Swift}) in the 
X-rays and UV and optical, the {\it Chandra} X-ray Observatory, using either HRC \citep{mur00} or
ACIS instruments \citep{nou98}.
Here we present a complete analysis of optical/UV and X-ray SSS phase of this nova exploiting
all the available data. Section \ref{data} describes the data reduction, Section \ref{anal} the analysis,
and in Section \ref{discussion} we discuss the results.

\section{Data reduction} \label{data}

An 24 ks {\it XMM-Newton} ToO was dedicated to M31N2013 on Feb. 8 (see Table 1).
We made use of SAS v.15 to perform a standard data reduction from Observation Data Files, 
followed by a barycentrization to the Solar System using the SAS tool {\tt barycen}.
For the timing analysis we only made the standard selection on pattern (0-4 for pn and 0-12 for MOSs), using the 0.2-1 keV 
energy band since the source is not detected above 1 keV. Pn and MOSs events were then added to perform the analysis.
We then made use of the EXTraS tools and cuts \footnote{http://www.extras-fp7.eu/index.php} to study M31N2013 light curves in the 0.2-1 keV 
energy band.
For the spectral analysis we used the entire EPIC data set, excluding high background periods (this reduced the exposure to $\sim$16 ks).
The Optical Monitor (OM)\citep{mas01} data were reprocessed using the standard SAS tool 
{\tt omichain}\footnote{http://xmm-tools.cosmos.esa.int/external/sas/current/doc/ }.
The transient was not detected by the Reflection Grating Spectrometers (RGS1,2) \citep{her01}.

The {\it Swift} XRT data of the 8 observations (see Table 1)  were processed with standard procedures using the 
Ftool (v.6.19) task {\tt xrtpipeline}. We selected events with grades
0-12 and limited the analysis between 0.3-1.0 keV.
The standard UVOT processing pipeline was used for the observations carried out with the UV filters,
UW1,UW2 and UM2. We used Heasoft tool {\tt uvotmaghist} (taking into account the Small Scale Sensitivity check as recommendend in the cookbook) to  
obtain a detection or an upper limit magnitude for each observation.
The {\it Chandra} observations (see Table 1), were reprocessed using the {\it Chandra Interactive Analysis of Observation }
(CIAO) software v4.5 with the {\tt chandra\_repro} task.\\
Table 1 reports the details of the X-ray and UV/optical simultaneous observations used in this paper.

\begin{table*}
\label{tab:tab-obs}
\begin{center}
  \caption{Summary of the X-ray and UV/optical simultaneous observations of M31N2013. 
 Here we report for each observation the telescope, observation id, the band in which optical/UV instrument was operating,
    the start time, the exposure time. We also report 
the X-ray equivalent luminosity (see Section \ref{anal}) and the average optical/UV magnitude for each observation. Errors are at 1$\sigma$,
  upper limits are at 3$\sigma$.}
\begin{tabular}{ccccccc}
\hline
Instrument & OBS-ID & Optical/UV Band & Tstart & Exposure & L$_{50}$ & Optical/UV magnitude\\
- & - & - & date(UTC) & ks & $10^{37}$ erg s$^{-1}$ & mag \\
\hline
{\it Swift} & 00032697001 & U & 2013-02-01 11:40:59 & 3.9 & $<$3.6 & 18.84$\pm$0.08\\
{\it Swift} & 00032697002 & U & 2013-02-05 19:48:59 & 3.5 & 3.5$\pm$1.3 & 19.85$\pm$0.15\\
{\it XMM-Newton} & 0701981201 & UVW1 & 2013-02-08 21:55:52 & 24.0 & 3.3$\pm$0.2 & 19.93$\pm$0.38\\
{\it Swift} & 00032697003 & U & 2013-02-09 13:42:58 & 3.9 & 1.8$\pm$0.8 & 20.37$\pm$0.18\\
{\it Swift} & 00032697004 & U & 2013-02-13 07:39:59 & 3.8 & $<$3.7 & 20.78$\pm$0.35\\
{\it Swift} & 00032697005 & U & 2013-02-17 04:27:59 & 4.0 & $<$3.5 & $>$21.9\\
{\it Chandra/ACIS-I} & 14930 & - & 2013-02-18 08:02:31 & 3.9 & $<$250 & -\\
{\it Swift} & 00032697006 & UW1 & 2013-02-24 01:14:13 & 6.0 & $<$2.6 & $>$22.7\\
{\it Chandra/HRC-I} & 14400 & - & 2013-02-24 10:38:42 & 5.1 & $<$0.94 & -\\
{\it Swift} & 00032697007 & UW2 & 2013-02-26 12:36:58 & 8.9 & $<$1.9 & $>$21.8\\
{\it Swift} & 00032697008 & UM2 & 2013-03-07 03:18:38 & 5.2 & $<$3.3 & $>$21.8\\
{\it Chandra/HRC-I} & 15620 & - & 2013-03-11 23:03:17 & 5.1 & $<$0.98 & -\\
{\it Chandra/ACIS-I} & 14931 & - & 2013-03-12 00:43:55 & 3.9 & $<$250 & -\\
\hline
\end{tabular}
\end{center}
\end{table*}

\section{Data analysis} \label{anal}

\subsection{X-ray data}

M31N2013 was selected among the 3XMM source list in M31 as one of the best eclipsing object candidate.
We adopted the spectral model used for the equivalent luminosity (see below) to convert EXTraS background-subtracted count rates into absorbed fluxes, 
then we made a weighted average of the flux light curve of each EPIC instrument to obtain a total flux light curve.
We fitted a periodic eclipse model on the filtered  500 s flux light curve.
Using an f-test, we found a significant improvement with respect to a
constant model, with a chance probability of 3.4$\times10^{-6}$. However, the model itself does not give a good representation
of the light curve shape, giving null hypothesis probability (nhp) of 1.2$\times10^{-4}$, indicating a more complex shape.
We performed an independent search on pn plus MOSs events with a Rayleigh test on the {\it XMM-Newton} pn plus MOSs event list \citep{buc83} over more than 9000 independent periods 
in the range 0.001-4.2 h.
This yielded a Z$^2$ value of 90 for P=1.29 h , which corresponds to a chance probability of 3$\times10^{-16}$ (taking into account the number of trials).
{We note that $\sim75$\% of events come from pn. A similar analysis based on pn only events still gives the periodicity at $>6\sigma$.}
To refine the period value we used a standard folding analysis based on the sum of pn and MOS counts into 10 phase bins. 
This resulted in a best period of P=1.28$\pm$0.02 h. All uncertainties are 1$\sigma$ error as 
estimated following \citet{lea87}.

We fitted the {\it XMM-Newton} spectra using an absorbed \citep[abundances from ][]{wil00} blackbody,
obtaining a good fit (nhp=0.16, d.o.f.=37) with column density 
N$_H$ = (2.6$\pm$0.4)$\times10^{21}$ cm$^{-2}$ (in this paper, all the uncertainties are at 1$\sigma$),
a temperature kT = (47$\pm$3) eV, an emission radius R = 4.2$_{-1.6}^{+2.9}\times 10^{9}$ cm
and an absorbed 0.2-10 keV flux of (6.8$\pm$2.5)$\times10^{-14}$ erg cm$^{-2}$ s$^{-1}$.
For such a supersoft source, degeneration among parameters is expected and the flux cannot be de-absorbed correctly.
In order to compare M31N2013 with the other novae from M31, we
adopted the definition of ``equivalent luminosity'' L$_{50}$ from \citet{hen14}, obtained using a blackbody temperature
kT = 50 eV and the Galactic foreground absorption N$_H$ = $6.7\times10^{20}$ cm$^{-2}$ \citep{hen14}.
For M31N2013 we therefore obtain an unabsorbed flux of (4.6$\pm$0.2)$\times10^{-13}$ erg cm$^{-2}$ s$^{-1}$, 
that translates into a
L$_{50}$ = (3.3$\pm$0.2)$\times10^{37}$ erg s$^{-1}$. See also Figure \ref{fig:lc-long}.

The {\it XMM-Newton} light curve of M31N2013 has been folded using 20 bins (so that each bins has $\sim$50 counts, 
see Section \ref{data}) using the period obtained through the timing analysis (see Figure 1).
The phased light curve shows a quasi-sinusoidal shape, even if the low significance of a constant plus sinusoidal model 
(nhp = 4.2$\times10^{-5}$, 17 d.o.f.)  and of a simple eclipse
model (nhp = 4.6$\times10^{-4}$, 16 d.o.f) point to a more complex behavior. 
Using a sinusoidal model, the pulsed fraction 
(defined as the normalization of the sinusoid over the constant) is (40$\pm$5)\%.
The light curve also shows hints of a linear decrease with time (see Figure \ref{fig:lc-xmm}), but not
statistically significant (an f-test gives a chance probability for the improvement of 0.07).

\begin{figure}[ht!]
\plotone{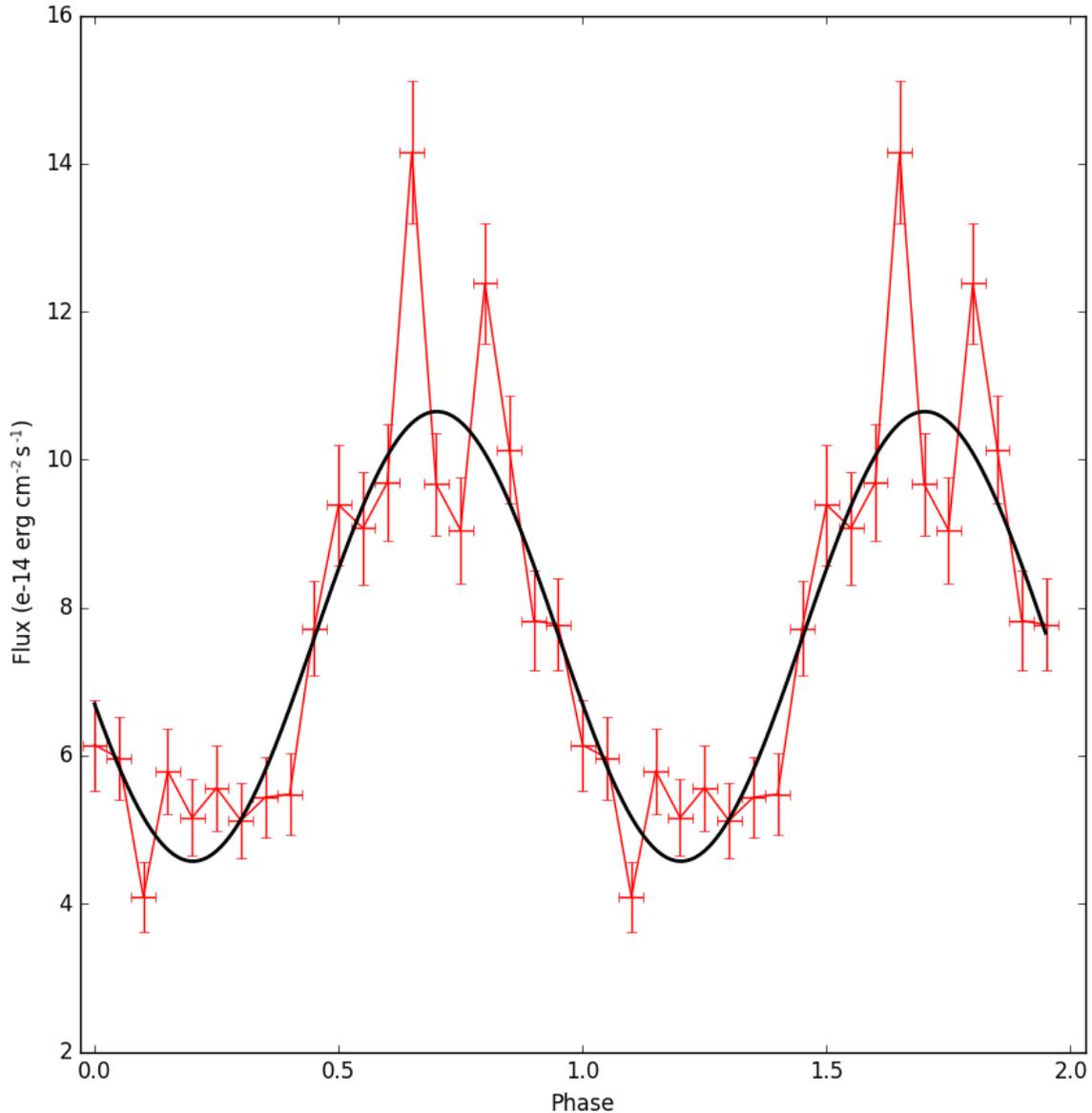}
\caption{{\it XMM-Newton} phased light curve in the 0.2-1.0 keV range adopting the best fit period of 
4595\,s. Count rates are converted to fluxes 
as detailed in Section \ref{anal}. Here, we also report a sinusoidal fit (nhp = 4.2$\times10^{-5}$ , 17 d.o.f.).
  Errors are at 1$\sigma$. 
  \label{fig:lc-phased}}
\end{figure}

\begin{figure}[ht!]
\plotone{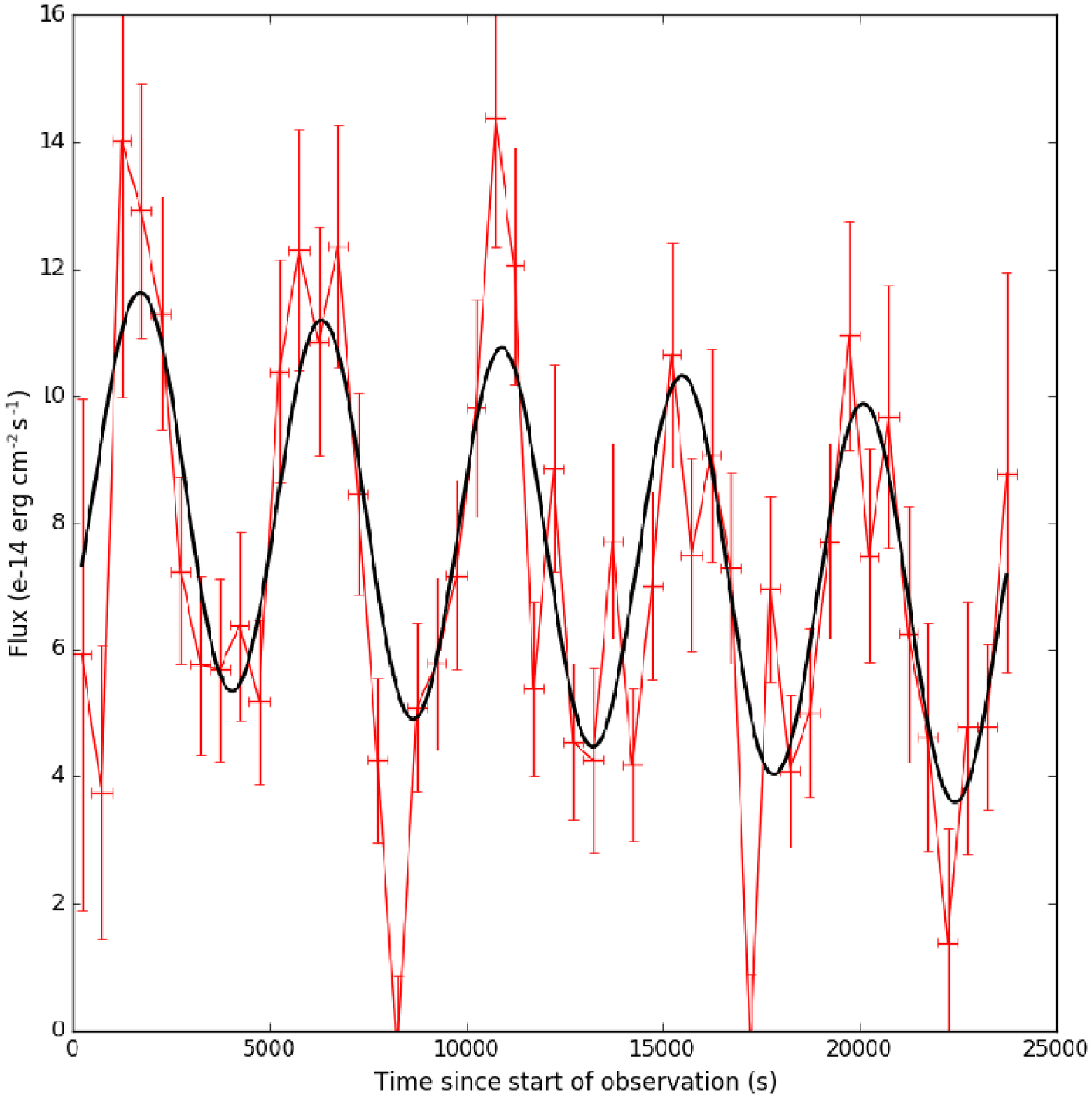}
\caption{Time resolved  {\it XMM-Newton} light curve in the 0.2-1.0 keV range along with a sinusoidal plus
linear fit (nhp = 1.2$\times10^{-4}$ , 47 d.o.f.). Errors are at 1$\sigma$ and bins are of 500 s. Fluxes are derived
as described in Section \ref{anal}. On the x axis, we report time since 2013-02-08 UTC 21:55:52.
  \label{fig:lc-xmm}}
\end{figure}

In order to search for spectral variability, we extracted {\it XMM-Newton} spectra in two phase ranges, 
namely 0.0-0.5 (off-pulse phase) and 0.5-1 (on-pulse phase).
A spectral fit with all the parameters tied but the normalization gives an acceptable fit, nhp=0.98 d.o.f.=30 (R$^{off}$=3.1$_{-2.4}^{+1.5}\times 10^{9}$ cm, 
R$^{on}$=4.1$_{-1.8}^{+3.3}\times 10^{9}$ cm). Similarly, leaving free the column density
we obtain an acceptable fit, nhp=0.96 d.o.f. 30 (N$_H^{off}$ = (2.4$\pm$0.5)$\times10^{21}$, N$_H^{on}$ = (1.8$\pm$0.4)$\times10^{21}$).
We also checked the spectra 
of the first and second half of observation, obtaining again an acceptable fit.
To further inspect possible spectral changes, we also produced Hardness Ratios (HR) both phased (10 bins) and 
with time (one per orbital period) in the energy bands
0.2-0.4 keV and 0.4-0.7 keV. The HR distributions are consistent with a constant model.
We therefore do not detect spectral variation with phase or time.

Using the {\tt ximage} {\tt detect} Heasoft tool, we checked for source detection in all the {\it Swift/XRT} 
observations. Obs.id. 00032697002 and 00032697003
revealed a source at the position of M31N2013 at 2.5$\sigma$. By adding the two images, we obtain a detection at 3.6$\sigma$. 
Altought statistically poor, 
we consider these as detections due to the positional coincidence and the fact that the 
{\it XMM-Newton} detection is in between these  two observations.    
The background-subtracted count rates were then converted into
equivalent luminosities L$_{50}$ of (3.5$\pm$1.3) and (1.8$\pm$0.8) $\times10^{37}$ erg s$^{-1}$
respectively, taking into account the exposure map, effective area and response matrix. For all the other {\it Swift/XRT} observations we extracted 
a 3$\sigma$ upper limit L$_{50}$ using the definition of
signal-to-noise. Similarly, we did not detect the X-ray counterpart in any of the {\it Chandra} data set, 
hence we derive 3$\sigma$ upper limits to  L$_{50}$ reported in Table 1 and Figure \ref{fig:lc-long}.

\subsection{UV and optical data}

An UV counterpart of M31N2013 is detected ($>3\sigma$) only during the first 4 {\it Swift/UVOT} observations and by the {\it XMM-Newton} OM
(see Table 1), when the nova was brighter. 
There is a clear exponential decrease in the U-band during the $\sim$12 days in which the source was detected.
For the first three {\it Swift/UVOT} observations we also obtained the UVOT positions overlapping at 1$\sigma$
and coincident with the optical position from \citet{hor13,sha13} 
(R.A.,DEC. J2000): $00^{{\rm h}}~44^{{\rm m}}~02.^{{\rm s}}09$, $+41^{\circ}~25'~44.29''$
($\pm$0.57'', systematic plus statistical error),
$00^{{\rm h}}~44^{{\rm m}}~02.^{{\rm s}}10$, $+41^{\circ}~25'~44.53''$ ($\pm$0.57'') and
 $00^{{\rm h}}~44^{{\rm m}}~02.^{{\rm s}}12$, $+41^{\circ}~25'~44.24''$ ($\pm$0.62'').\\
To construct a long-term UV/optical light curve of M31N2013 we made use of the
measurements reported in Table 2, {\it Swift/UVOT} and {\it XMM-Newton OM} data in Table 1.
The long-term UV/optical light curve with all collected photometry is shown in the upper panel of 
Figure\,\ref{longterm} along the X-ray equivalent luminosities L$_{50}$ from 
{\it XMM-Newton}, {\it Swift} and {\it Chandra}, reported in the lower panel.
Bearing in mind that novae show different color decays and brightness at different wavelengths and we are using measurement from different filters, we
estimated $t_2$ using a power law fit to the composite long-term light curve. We derive $\alpha$=1.7 and $t_2$=2.8$\pm$0.2 d
which is consistent with the simple difference between the two R band measures (3.0 days).

\begin{table*}
\label{tab:tab-obs2}
\begin{center}
  \caption{Summary of optical observations of M31N2013.W refers to white (broadband) filter. CBAT data of Jan. 27 do not report the error: we assumed the same error as the other measurement,
      so that it should be taken with caution.
    References are: [1] \citep{hor13}; [2] \citep{sha13}; [3] IAU CBAT\footnote{http://www.cbat..eps.harvard.edu/unconf/followups/J00440207+4125442.html}
}

\begin{tabular}{cccc}
\hline
Time & Optical band & magnitude & reference\\  
UTC & - & mag & -\\
\hline
2013-01-12 16:52:19.2 & R & $>$20.5 & [1]\\
2013-01-20 10:06:14.4 & W & $>$19.1 & [3]\\
2013-01-23 10:06:10.0 & W & $>$19.1 & [3]\\
2013-01-25 17:28:19.2 & R & 15.05$\pm$0.07 & [1]\\
2013-01-26 10:14:52.8 & W & 15.6$\pm$0.2 & [3]\\
2013-01-27 10:42:14.4 & W & 15.7$\pm$0.2 & [3]\\
2013-01-28 01:44:49.9 & U & 17.4$\pm$0.2 & [3]\\
2013-01-28 17:26:52.8 & R & 17.02$\pm$0.09 & [2]\\
\hline
\end{tabular}
\end{center}
\end{table*}

\section{Discussion} \label{discussion}

\subsection{The X-ray periodicity}

Our analysis of the {\it XMM-Newton} X-ray data resulted in a periodic modulation with a period
of 1.28$\pm$0.02 h, obtained in an observation lasting about 5\,cycles. 
The amplitude of this variability is large $\sim$40$\%$.
The modulation appears to be also structured with hints of a double-peaked maximum and a flat minimum. 
The timescale and large amplitude of the modulation exclude that the SSS luminosity variation is due to pulsations
(see Section \ref{sec:intro}). We ascribe the periodicity to the orbital period of the system.
It is the first nova in M31 with an orbital period unambiguously detected in the X-rays.
Only candidate periods of two novae in M31 
were found by \citet{hen10,hen14} during the SSS phase of M31N 2006-04a and M31N 2011-11e, 
1.6\,h and 1.3\,h, respectively.
The orbital period of CVs is an observational proxy  
 of their evolutionary state, where the secular mass transfer rate decreases
as the systems evolve
towards short orbital periods \citep{how01,bar03}.
Hence, systems below the 2-3\,h orbital period gap \citep{war95}
are old systems and expected to accrete at a low-rates 
$\lesssim 5\times 10^{-11}\, M_{\odot}\,yr^{-1}$ \citep{kni11}.
Assuming a 1.3\,h orbital period, the donor in M31N2013 is expected to be a very low-mass ($\rm M_2 \leq 0.1 M_{\odot}$)
star and of late M or even later spectral type \citep{kni11}. Observed spectral 
types of CV donors at these short periods have been found to be late M dwarfs,  M6-M9 as it is the
case for the short period (1.4\,h) old novae CP\,Pup or GQ Mus ($>$M6) \citep{szk88}.
For L-type donors the systems could also be period-bouncers 
and a few have been found so far with only one confirmed X-ray emitting magnetic system
\citep{ste17} accreting at very low rate ($\sim 10^{-14}\,M_{\odot}\,yr^{-1}$).
The very short 1.28\,h  period locates M31N 2013-01b close to the expected theoretical orbital minimum of CVs ( 1.1\,h) 
\citep{how01}.  
Comparing the orbital period distribution of novae from \citet{rit03} this nova is among the shortest orbital period systems
and one of the few detected to display an orbital modulation in the SSS phase.

Orbital X-ray variability could be due to structures in the accretion disc although  absorption effects should be
present. Additionally, a partial eclipse from fixed regions such as the disc-rim or by the donor star could be
viable solutions.
Similar partial eclipses were claimed for two Galactic novae HV Cet \citep{bea12} and
V5116\,Sgr \citep{sal08}. Both the models require an high-inclination system (i$\gtrsim60^{\circ}$)
Another  possibility is that M31N2013 harbours a magnetic WD of the polar type
\citep[see ][ for a review]{cro90}, where no disc is formed due to the high 
magnetic field of the WD primary
($\rm B \geq 10\,MG$) that locks its rotation at the orbital period. The polars show strong X-ray variations
(up to 100$\%$) due to localised accretion spots   at the magnetic poles. However, the burning
in the very early phases of a nova would rapidly reach a  spherical symmetry, heating the whole WD surface
\citep{cas10}. The radius of the SSS emitting region is found to be in the range 
$\sim 3-7\times 10^{9}$\,cm that is indeed much larger than those of WDs, indicating an expanded emitting envelope.
The lack of observed spectral variability along the phase may favour an eclipse, possibly 
partial, of the X-ray emitting region.

\subsection{M31N\,2013-Ib: a very fast nova in M31}

M31N2013 is a very fast nova, with a $t_2$ decay in the R band of $\sim$3\,d.
The rate of decay is then $d\,m/d\,t \sim 0.7$ and consistent with the universal law by \citet{hac06}
where the flux decays as $\rm F \propto t^{-\alpha}$ with $\alpha$=1.7-1.75. 
We find $\alpha$=1.7.
\citet{lee12} suggests that such very fast novae are more rarely encountered in M31 than in our Galaxy, even if selection effects should be taken into account.
In particular, the online catalogue of optical apparent novae in M31 by Pietsch \& Haberl\footnote{http://www.mpe.mpg.de/~m31novae/opt/m31/index.php}
reports a handful (11) of very fast novae with similar $t_2\sim$2-4\,d. 
Fast novae are found to harbour massive WD ($\rm M_{WD} \geq 1.2\,M_{\odot}$) \citep{del95,del02}
and  the light curves are predicted to be almost independent of the chemical composition once iron content is
fixed \citep{hac06}. 

The SSS X-ray turn-on and turn-off times cannot be fully constrained due to the lack of a deep monitoring 
shortly after the optical maximum. While the {\it Swift/XRT} upper limit on Feb. 1 is too weak to constrain the turn-on time,
the detections by {\it Swift}/XRT and {\it XMM-Newton} could hint to a fast turn-on t$_{on}<$ 10\,d (taking into account the T0 reported in Section \ref{sec:intro}).
The turn-off time could have occurred about one month 
after the optical maximum given the non-detection in the {\it Chandra} data, or before. Such short times are consistent
with the correlation found between the turn-on and turn-off times found for M31 novae by \citet{hen14}.
The hot blackbody temperature as derived from the X-ray spectral fit is also in agreement with a short turn-off
time on similar timescale \citep[see e.g. ][]{hen14}.
Additionally, results from the M31N2013 datasets -- including the short t2, the extremely high expension velocity
detected in the optical spectrum of the early phases of the outburst \citep{sha13}, turn-on and turn-off time -- appear to be
consistent with the relations found for the M31 novae \citep{hen14}.
Althought consistent, in all these graphs M31N2013 is always at the edge of the relationships, owing to 
its short t2, turn-on time and turn-off time, thus allowing for a better constraint of the link among these quantities.
The X-ray luminosity of the SSS of M31N2013 is consistent with other M31 novae found 
so far (L$_{50}<8.7\times10^{37}$ erg s$^{-1}$, \cite{hen14}), although it is in the bulk
of the most luminous ones (only 4 out of the 24 M31 novae reported in \citet{hen14} show a higher L$_{50}$).

Using the dust maps by \citet{mon09} that include the contribution of the Galactic foreground extinction 
(E$_{gal}$(B-V)=0.10), 
 we derive a total extinction $\rm 0.10\lesssim E(B-V)\lesssim 0.21$, since the location of M31N2013 within  M31 is not known.
This translates into a range of extinction in the R band of $\rm 0.26 \leq A_R \leq 0.55$.
Assuming a distance of 776$\pm$18\,kpc and this range of extinction,
we estimate an absolute magnitude at the peak of the optical light curve 
in the range $\rm -9.60 \leq M_R \leq -10.0$.  This range is fully consistent with the absolute peak magnitude
observed in fast novae with similar $t_2$ and in particular with the super-bright Galactic nova
V1500\,Cyg \citep{kat13}. Unfortunately, due to the loosly constrained optical and 
X-ray positions, a search for the nova progenitor resulted inconclusive.

Although not studied until now, M31N2013 is one of the brightest and fastest nova ever detected in  M31, belonging to the
disc population and likely harbouring a massive WD and one of the few known at very short orbital periods.

\begin{figure}[ht!]
\plotone{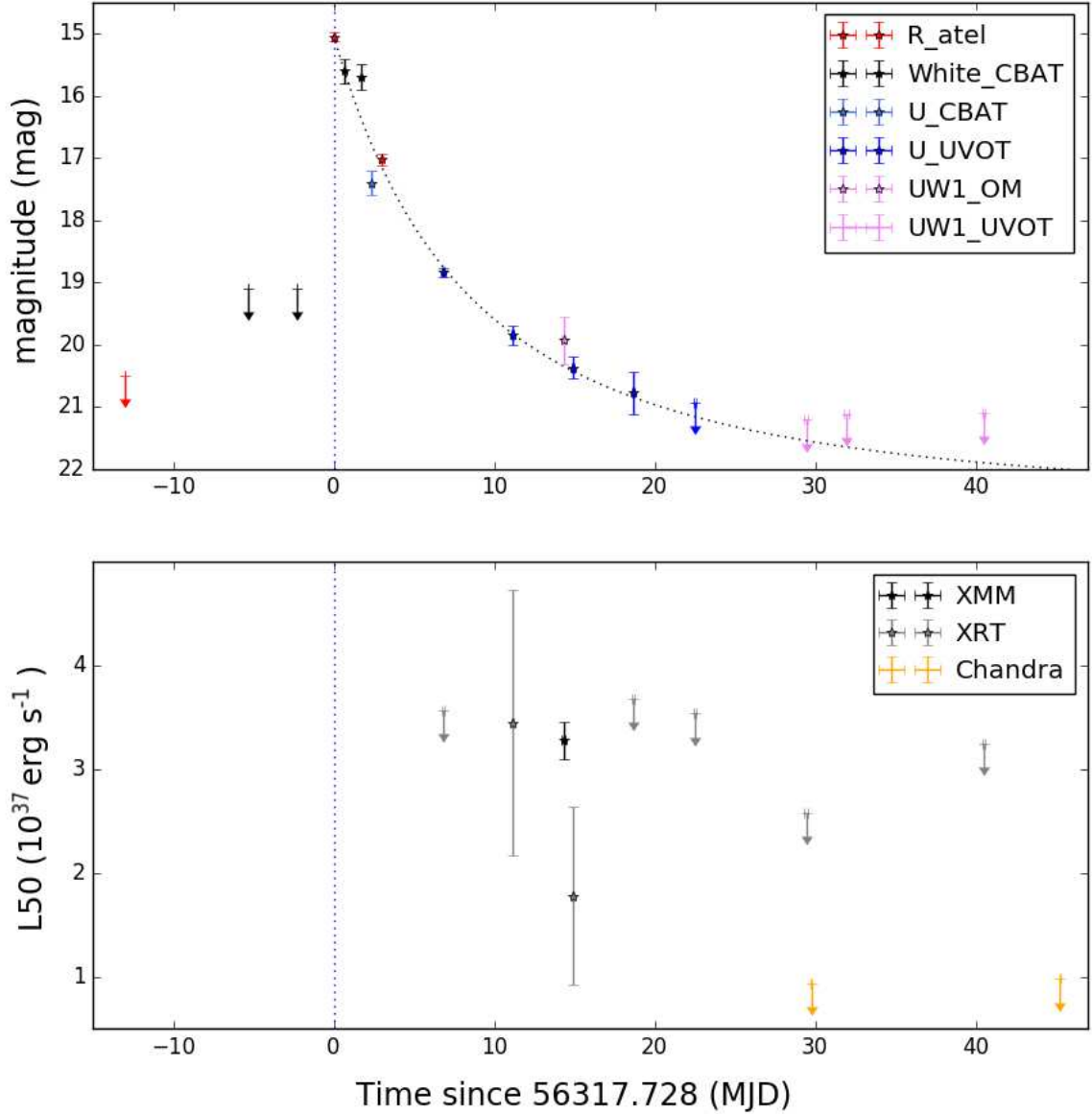}
\caption{The long term light curves of M31N2013. On the x axis, we report time since the first detection of Nova. {\it Upper Panel:} the UV/optical magnitudes  clearly show an
exponential decrease of the optical nova (dashed line) after the maximum (vertical dotted line). 
{\it Lower Panel:} the equivalent X-ray luminosity L$_{50}$, as obtained from all X-ray instruments, shows instead 
a less clear decay of the X-ray source. Errors are at 1$\sigma$ and upper limits at 3$\sigma$.
  \label{fig:lc-long}}
\label{longterm}
\end{figure}

\acknowledgments

The authors acknowledge financial support from ASI under ASI/INAF agreement N.2017-14.H.0.
This work was supported by the Fermi contract ASI-INAF I-005-12-0. 

\vspace{5mm}
\facilities{XMM,CXO,Swift}


\begin{thebibliography}{}
\bibitem[Barker\&Kolb(2003)]{bar03} Barker, J. \& Kolb, U., 2003, MNRAS, 340, 623
\bibitem[Beardmore et al.(2008)]{bea08} Beardmore, A.P. et al., 2008, ASPC, 401, 296
\bibitem[Beardmore et al.(2012)]{bea12} Beardmore, A.P. et al., 2012, A\&A, 545, 116
\bibitem[Bode\&Evans(2008)]{bod08} Bode, M.F \& Evans, A., 2008, Classical Novae, eds. M.F. Bode, \& A. Evans (Cambridge University Press)
\bibitem[Buccheri et al.(1983)]{buc83} Buccheri, R., et al., 1983, A\&A, 128, 245
\bibitem[Casanova et al.(2010)]{cas10} Casanova, J., et al., 2010, A\&A, 513, L5
\bibitem[Cropper(1990)]{cro90} Cropper, M., 1990, SSRv, 54, 195
\bibitem[Dalcanton et al.(2012)]{dal12}  Dalcanton, J. J., et al., 2012, ApJS, 200, 18
\bibitem[De Luca et al.(2016)]{del16} De Luca, A., Salvaterra, R., Tiengo, A., et al. 2016, in The Universe of Digital Sky Surveys, Astrophysics and Space Science Proceedings, Vol. 42, ed. N. R. Napolitano et al. (Cham: Springer International), 291
\bibitem[Della Valle\&Livio(1995)]{del95} Della Valle, M. \& Livio, M., 1995, ApJ, 452, 704
\bibitem[Della Valle et al.(2002)]{del02}  Della Valle, M. et al, 2002, A\&A, 390, 155
\bibitem[Den Herder(2001)]{her01} Den Herder, J.W., et al. 2001, A\&A 365, L7
\bibitem[Dobrotka\&Ness(2010)]{dob10} Dobrotka, A. \& Ness, J.-U., 2010, MNRAS, 405, 2668
\bibitem[Drake et al.(2003)]{dra03} Drake, J. J. et al., 2003, ApJ, 584, 448
\bibitem[Hachisu\&Kato(2006)]{hac06} Hachisu, I. \& Kato, M., 2006, ApJS, 167, 59
\bibitem[Henze et al.(2010)]{hen10} Henze, M., et al., 2010, A\&A, 523, 89
\bibitem[Henze et al.(2014)]{hen14} Henze, M.  et al. 2014, A\&A 563, A2
\bibitem[Hornoch(2013)]{hor13} Hornoch, K., 2013, ATel.4765
\bibitem[Howell et al.(2001)]{how01} Howell, S.B., et al., 2001, ApJ, 550, 897
\bibitem[Kato et al.(2013)]{kat13} Kato, M. et al., 2013, ApJ, 779, 19
\bibitem[Knigge et al.(2011)]{kni11} Knigge, C., Baraffe I. \& Patterson, J., 2011, ApJS, 194, 28
\bibitem[Krautter(2002)]{kra02} Krautter, J., 2002 in Classical Nova Explosions, eds. M. Hernanz, \& J.Jose AIP Conf. Proc., 637, 345
\bibitem[Leahy(1987)]{lea87} Leahy, D. A., 1987, A\&A, 180, 275
\bibitem[Lee et al.(2012)]{lee12} Lee C.-H., et al., 2012, A\&A, 537, A43
\bibitem[Mason et al.(2001)]{mas01} Mason K.O., et al. 2001, A\&A 365, L36
\bibitem[Montalto et al.(2009)]{mon09}  Montalto, M. et al., 2009, A\&A, 507, 283
\bibitem[Murray et al.(2000)]{mur00} Murray, S.S., et al., 2000, Proc.SPIE, 4012, 68
\bibitem[Ness et al.(2015)]{nes15} Ness, J.-U. et al., 2015, A\&A, 578, 39
\bibitem[Nousek et al.(1998)]{nou98} Nousek, J.A., et al., 1998, Proc.SPIE, 3444, 225
\bibitem[Orio et al.(2001)]{ori01} Orio, M., et al., 2001, A\&A, 373, 542
\bibitem[Page et al.(2013)]{pag10} Page et al., 2013, ApJ, 768, L26
\bibitem[Payne-Gaposchkin(1964)]{pay64} Payne-Gaposchkin, C., 1964, The galactic novae (New York: Dover Publication)
\bibitem[Pietsch et al.(2011)]{pie11} Pietsch, W. et al. 2011, A\&A 531, A22
\bibitem[Ritter\&Kolb(2003)]{rit03} Ritter H. \& Kolb U., 2003, A\&A, 404, 301
\bibitem[Sala et al.(2008)]{sal08} Sala, G., et al., 2008, ApJ, 675, L93
\bibitem[Schwarz et al.(2011)]{sch11} Schwarz, G.J., et al., 2011, ApJS, 197, 31
\bibitem[Shafter et al.(2011)]{sha11} Shafter, A.W., et al., 2011, ApJ, 734, 12
\bibitem[Shafter et al.(2013)]{sha13} Shafter, A.W., 2013, ATel.4768
\bibitem[Shore et al.(2011)]{sho11} Shore, S.N., et al., 2011, A\&A, 533, 8
\bibitem[Stelzer et al.(2017)]{ste17} Stelzer, B., et al., 2017, A\&A, 598, L6
\bibitem[Strope et al.(2010)]{str10} Strope, R.J., et al. 2010, AJ, 140, 34
\bibitem[Str\"uder et al.(2001)]{str01} Str\"uder, L., et al. 2001, A\&A 365, L18
\bibitem[Szkody\&Feinswog(1988)]{szk88} Szkody \& Feinswog, 1988, ApJ, 334, 422 
\bibitem[Turner et al.(2001)]{tur01} Turner, M.J.L., et al., 2001, A\&A 365, L27
\bibitem[Warner(1995)]{war95} Warner, B., 1995, Camb. Astrophys. Ser., Vol. 28
\bibitem[Wilms et al. (2000)]{wil00} Wilms, J., Allen, A. \& McCray, R.\ 2000, ApJ, 542, 914
\bibitem[Wolf et al. (2018)]{wol18} Wolf, W.M., et al. 2018, ApJ 855, 127
\end{thebibliography}
\end{document}